\documentclass{article}
\usepackage{ijcai24}

\usepackage{times}
\usepackage{soul}
\usepackage{url}
\usepackage[hidelinks]{hyperref}
\usepackage[utf8]{inputenc}
\usepackage[small]{caption}
\usepackage{graphicx}
\usepackage{amsmath}
\usepackage{amsthm}
\usepackage{amssymb}
\usepackage{booktabs}
\usepackage{algorithm}
\usepackage{algorithmic}
\usepackage[switch]{lineno}
\usepackage[table,dvipsnames]{xcolor}
\usepackage{tabularx}
\usepackage{listings}
\usepackage{longtable}
\usepackage{threeparttable}
\usepackage{textcomp}
\usepackage{multirow}
\usepackage{multicol} 
\usepackage{verbatim}
\usepackage{booktabs}
\usepackage{enumitem}
\usepackage{listings}
\usepackage{mathtools}
\usepackage{pifont}
\usepackage[most]{tcolorbox}
\usepackage{pifont}
\usepackage{makecell}
\usepackage{appendix}

\usepackage{graphicx}
\usepackage{float}
\usepackage{subfig}

\usepackage{twemojis}
\usepackage{diagbox}
\usepackage{threeparttable}

\usepackage[square]{natbib}
\defcitealias{Fair_use}{Copyright Office, 2023}
\defcitealias{Central_Cyberspace_Affairs_Commission_Office}{CAC, 2023}
\defcitealias{Midjourney}{Mijourney Inc., 2023}
\defcitealias{copyleft_definition}{FSF, 2023}
\defcitealias{GPL}{FSF, 2007}
\defcitealias{Permissive}{FreeBSD, 2024}
\defcitealias{di2010exploratory}{Di Penta et al., 2010}
\defcitealias{complaint}{New York Times, 2023}
\defcitealias{Michael_2023}{Grynbaum and Mac, 2023}

\newcommand{\tabitem}{~~\llap{\textbullet}~~}

\newcommand{\todo}[1]{\textcolor{blue}{[\textbf{TODO!} #1]}}

\usepackage{xurl}

\newtcolorbox{boxK}[2][]{
    left=1mm, right=1mm, bottom=1mm,
    colframe=RoyalPurple!55!Aquamarine!100!,
    fonttitle = \bfseries,
    colbacktitle=RoyalPurple!55!Aquamarine!100!,
    squeezed title= #2,#1,
    attach boxed title to top left={yshift=-3mm, xshift=0.0cm},
    colback=SeaGreen!10!CornflowerBlue!10,
    toprule = 4pt, % top rule weight
    enhanced,
    width=1.0\linewidth,
    fuzzy shadow = {0pt}{-2pt}{-0.5pt}{0.5pt}{black!35} 
}

\newcounter{minilabel}
\newcommand{\minilabel}[1]{%
  \refstepcounter{minilabel}%
  \addcontentsline{rlo}{section}{\protect\numberline{\theminilabel}#1}%
}

\title{Copyleft for Alleviating AIGC Copyright Dilemma:  What-if Analysis, Public Perception and Implications}
\author{
Xinwei Guo$^{1}$\footnotemark[1]
\and
Yujun Li$^{2}$\footnotemark[1]
\and
Yafeng Peng$^{2}$\And
Xuetao Wei$^{1}$
\\
\affiliations
$^1$Southern University of Science and Technology\\
$^2$Jiangxi Normal University\\
\emails
guoxw2023@mail.sustech.edu.cn,
\{liyujun, yafeng\}@jxnu.edu.cn,
weixt@sustech.edu.cn
}

\begin{document}

\maketitle

\renewcommand{\thefootnote}{\fnsymbol{footnote}}
\footnotetext[1]{These authors have contributed equally to this work.}
% \footnotetext[2]{Corresponding author}

\begin{abstract}
%At present, AIGC is faced with various copyright problems, which may make the management of AIGC more strict and hinder the development of AIGC. Therefore, this paper aims to explore the advantages and limitations of a new solution Copyleft applied to the field of AIGC, and finally analyze the applicability of copyleft in AIGC.

As AIGC has impacted our society profoundly in the past years, ethical issues have received tremendous attention. The most urgent one is the AIGC copyright dilemma, which can immensely stifle the development of AIGC and greatly cost the entire society. Given the complexity of AIGC copyright governance and the fact that no perfect solution currently exists, previous work advocated copyleft on AI governance but without substantive analysis. In this paper, we take a step further to explore the feasibility of copyleft to alleviate the AIGC copyright dilemma. We conduct a mixed-methods study from two aspects: qualitatively, we use a formal what-if analysis to clarify the dilemma and provide case studies to show the feasibility of copyleft; quantitatively, we perform a carefully designed survey to find out how the public feels about copylefting AIGC. The key findings include: a) people generally perceive the dilemma, b) they prefer to use authorized AIGC under loose restriction, and c) they are positive to copyleft in AIGC and willing to use it in the future.

\end{abstract}
\section{Introduction}

With the advent of the large model era, AIGC (Artificial Intelligence Generated Content) applications such as Midjourney\citepalias{Midjourney} and GPT series\citep{GPT_series} have demonstrated impressive performance, marking the AIGC technology entering a new historical stage\citep{cao2023comprehensive}.
Literally, AIGC refers to content generated by artificial intelligence models, which promises to replace a large amount of work automatically, thus significantly improving users' productivity. Because of the powerful capabilities and broad prospects demonstrated by AIGC technologies, it has received widespread attention from society\citep{zhang2023complete, mesko2023imperative}. However, due to the lack of sound regulations for AIGC, it faces various burning ethical issues, including copyright disputes.

For example, in Case 1, OpenAI was sued by the New York Times for illegally using unauthorized copyright content in model training.
In Case 2, the court did not support AIGC's copyright claim. Contrary to the verdict in Case 2, as Case 3 illustrates, disagreement remains on whether AIGC should be protected by copyright law.
Thus, settling AIGC copyright disputes is critical to advance its development and regulation\citep{ijcai2023p803,zhang2023navigating}.

\begin{boxK}{Case 1: The New York Times sued OpenAI}
    \small 
    % \textbf{New York Times sues openAI for the content infringement:}
    In December 2023, the New York Times sued OpenAI over copyright infringement, alleging OpenAI used the newspaper's material without permission to train the massively popular GPT\citepalias{Michael_2023,complaint}.
\end{boxK}

\begin{boxK}{Case 2: Unsupported AIGC copyright claim}
    \small 
    % AIGC copyright claim NOT supported
    % \textbf{A case that did NOT support the copyright claim:} \\
    In September 2022, Kris Kashtanova applied for copyright protection for her comic book \textit{Zarya of the Dawn} but did not reveal that the images used in the book were created by the AIGC tool, Midjourney. The US Copyright Office said that while the content of the book is protected by copyright, the copyright of AI-generated images is not protected\citep{Harvard}.
\end{boxK}

\begin{boxK}{Case 3: Supported AIGC copyright claim}
    \small 
    % \textbf{A case that supported the copyright claim\citep{Xie}:} \\
    In February 2023, the plaintiff generated a series of images using Stable Diffusion and uploaded them to a social platform.
    Then, the defendant removed the watermark embedded by the plaintiff and used them in his article.
    As China’s first lawsuit involving the copyright of AIGC,
    the plaintiff sued the defendant for infringement of his copyright, and the Beijing Internet Court supported the plaintiff's copyright claim in the first-instance judgement\citep{Xie}.
    
\end{boxK}

%\subsubsection{The AIGC Copyright Dilemma}

%After the quantity of the training set is basically satisfied, the quality is one of the most critical factors for the performance of a large model.During the barbaric growth of the AIGC, the difficulty in data collection made AI companies more inclined to use high-quality, well-typeset data sets such as newspapers and books, even if there is a severe risk of copyright infringement\citep{zhang2023navigating}. This directly pushes AIGC into the dilemma of copyright disputes, and may suffer a huge blow.

As reflected in these three cases, AIGC's copyright disputes involve three stakeholders, namely, (1) Owners of training data(referred to as \textbf{Data Owner}) whose data is utilized for training the AIGC models; (2) Owners of AIGC models (referred to as \textbf{Model Owner}) who develop the AIGC model and provide services; (3) Users of AIGC models (referred to as \textbf{Model User}) who develop the prompts, feed into AIGC models and generate the content. 

\begin{comment}
    
\begin{figure*}[ht]
\centering
% 如果introduction需要加文字，figure1可以缩到0.9
\includegraphics[width=0.9\linewidth, trim=1cm 1.0cm 1cm 1cm, clip]{images/figure1_4.jpg}
\caption{(a) The dilemma of AIGC under copyright; (b) The situation of AIGC under copyleft.
Hollow arrows and crosses of the same color indicate granting the copyright/copyleft of the work to a certain stakeholder and corresponding result, respectively.
For example, the red hollow arrow indicates granting copyright to the data owners, and the red cross indicates that the granting will hinder the development of step\ding{192} in the AIGC life cycle.
}
\minilabel{localref}\label{fig:1}

\end{figure*}
\end{comment}

The core question centered around these copyright disputes is: who owns the copyright of the content generated by artificial intelligence? 
The answer could be any stakeholder. This is a tricky problem as no matter which stakeholder obtains the copyright of AIGC, it will inevitably hinder the development of the AIGC ecosystem, which is referred to as the \textbf{AIGC copyright dilemma}. On the one hand, legislation for AIGC is a long and complicated process. On the other hand, the development of AIGC is of great significance for improving the productivity of the whole society. Solving this problem at the expense of AIGC development is not a wise solution. Therefore, a better solution is expected, one that can balance stakeholders' interests to resolve copyright disputes. 

Interestingly, previous work has its voice in advocating for copyleft in AI governance. Copyleft is a legal technique that has been practiced in GNU licenses for free software\citepalias{copyleft_definition}. It grants certain freedoms over copies of copyrighted works while requiring that the same rights be retained in derivatives. The ``contagiousness" of copyleft ensures that the work must be used under copyleft terms during dissemination. This restriction could promote the spread of AIGC while protecting copyright. \textbf{However, these previous work just mentioned the copyleft and lack a substantive analysis of its feasibility.} 

In this paper, \textbf{we take a step further to substantiate this advocate by answering the following three questions:}
\begin{itemize}
    \item What are the intricacies of the AIGC copyright dilemma?
    \item What is the public perception of the dilemma and copyleft?
    \item What are the implications and recommendations of these studies? 
\end{itemize}

First, we conduct \textbf{a formal and thorough what-if analysis to clarify the dilemma}. By assembling three scenarios of what if one stakeholder claims the copyright and what impact the other two stakeholders have, we progressively present the outcomes from three aspects: benefits, challenges, and risks. The results of the analysis clearly reveal the intricacies of this dilemma and show that no matter which stakeholder claims the copyright, the development of the other two stakeholders is inevitably hindered, which is not beneficial to the whole AIGC lifecycle. Given that there is no perfect solution to this dilemma, we must seek a ``trade-off" solution, which is also the reason why copyleft was advocated. It alleviates conflicts among stakeholders from the development point of view. We also conduct a qualitative analysis of three cases of copyright disputes with the application of copyleft and find that it could alleviate the conflicts and resolve the disputes. Furthermore, \textbf{we carefully design a rigorous survey to quantitatively study the public perception of such a dilemma and the extent to which people accept copyleft as a feasible solution.} Based on the survey data, we build a regression model to capture how various factors impact people's willingness to adopt copyleft in the future. The results overall support copyleft as a workable solution. 

\textbf{Our contribution can be summarized as follows:}
\begin{itemize}
    %\item We propose to adopt the copyleft to alleviate the AIGC copyright dilemma.
    \item We clarify the AIGC copyright dilemma via a formal and thorough what-if analysis and reveal the dilemma's intricacies, which serve as the ground for why copyleft was advocated.
    \item We conduct a quantitative analysis through a rigorous questionnaire survey to further assess the effectiveness of copyleft from the public perception perspective. 
    \item 
    The implications of our study show that (1) copyleft has the potential to mitigate stakeholders' conflicts and facilitate the development of AIGC; (2) The public clearly perceives the AIGC copyright dilemma and is generally positive about copylefting AIGC, thus open to the implementation of copyleft in AIGC. 
    % \todo{add implications we have learned from both qualitative and quantitative analysis} 
    
\end{itemize}

\textbf{Note that}, we admit the copyleft has its own limitations, e.g., privatization difficulty, which is also the reason we emphasize ``alleviate" rather than ``solve". Given the complexity of the AIGC copyright dilemma, we argue that it is better to find \textbf{a workable solution} that minimizes the risks of conflicts and promotes the significance of AIGC-empowered society development. This is also the major motivation for us to come up with this substantive analysis of the copyleft feasibility, which paves the way for future research and public discussion. We envision our work could be one of the beneficial supplements to AIGC copyright governance.   

%The remainder of this paper is organized as follows, section 2 reviews the related work over copyright issue in AIGC, section 3 gives an in-depth what-if-analysis of the AIGC copyright dilemma, section 4 introduces copyleft as a feasible solution, sections 5-6 presents a survey regarding public's perspective on copyright issue in AIGC and whether copyleft is a feasible solution. We conclude our paper with discussion, limitation, and conclusion (sections 7-9).

\section{Related Work}
The AIGC copyright dilemma did not arise unconsciously. 
In 2020, the European Union published a report on the challenges of AI to the intellectual property rights framework, which has already pointed out the copyright issues associated with AI-assisted outputs\citep{EU_2020}.
Shortly after that, the UK Intellectual Property Office launched an open consultation on copyright protection for computer-generated works\citep{kretschmer2022artificial}.
Over the last year, AIGC’s copyright issues have become a common concern; the US Copyright Office released guidance\citep{Federal_Register} on copyright registration of works containing AI-generated content,
and China issued interim measures\citepalias{Central_Cyberspace_Affairs_Commission_Office} to guide the development of AIGC.
Despite these efforts, the policies and guidelines are vague and difficult to materialize, and there are still many contentious lawsuits.

% AIGC-related lawsuits have challenged traditional
% copyright laws and triggered much discussion.
% From the perspective of the attribution problem, \citep{kahveci2023} analyzed the core difficulty in the verdict on the AIGC attribution: whether to accept the fair use argument\citepalias{Fair_use}. 
% %a doctrine in United States law that allows limited use of copyrighted material without first acquiring authorization. 
% Similarly, Alice et al.\citep{poland2023generative} discussed the potential harms of generative AI and different types of circumvention methods. 
% For the regulation of large generative AI models, \citep{hacker2023} believed that the disclosure of copyrighted material contained in training data could help authors enforce their rights. 
% Although these discussions analyzed the training set infringement in the training of AIGC model, they ignored AIGC's copyright protection and disputes.

In the current AIGC copyright protection and traceability research, watermarking is the mainstream method.
There are corresponding watermark technologies for different forms of generated content, whether it is image\citep{Fernandez_2023_ICCV,zhang2023editguard}, text\citep{kirchenbauer2023watermark} or audio\citep{Google_DeepMind_2023}.
In addition, there are also methods such as blockchain\citep{liu2023blockchain}, perturbation
\citep{chen2023challenges} and C2PA\citep{openai,c2pa} used by OpenAI in the DALL·E 3.
However, these methods are all aimed at solving the AIGC's accountability problem and cannot solve or alleviate the existing AIGC copyright dilemma.

In existing research on AI and licenses, \citep{contractor2022behavioral} used behavioral licenses to constrain downstream tasks and implement responsible AI.
However, the application object of their method is software or code, which does not involve copyright disputes as complex as those surrounding AIGC. Similarly, Schmit et al. \citep{schmit2023leveraging} simply mentioned that the copyleft can be used to achieve AI ethical governance in a three-page policy forum essay published online.
% \citep{schmit2023leveraging} simply mentioned the copyleft that can be used to achieve AI ethical governance in a two-page policy forum issue published online. 
However, they did not provide a detailed qualitative and quantitative analysis of its feasibility like ours.

\begin{table*}[ht]
    \centering
    \resizebox{\linewidth}{!}{
    \begin{tabular}{l|lll}
        \toprule
        Copyright Owner & Benefits     & Challenges       & Risks \\
        \midrule
        Data Owner      
        & \makecell[l]{
            \tabitem Protect the training data
        }   
        & \makecell[l]{
            \tabitem Huge scale \\
            \tabitem Diverse data sources \\
            \tabitem Hard to trace identities
        }
        & \makecell[l]{
            \tabitem Massive copyright lawsuits that hinder\\ the development of AIGC.
        }            
        \\
        \midrule
        
        % \rowcolor{gray!20}
        Model Owner  
        & \makecell[l]{
               \tabitem More revenue to develop better AIGC models\\ 
        }
        & \makecell[l]{
            \tabitem Trained models based on data that\\ \quad may not have copyright\\  
            \tabitem Rely on the user-constructed prompts
        }   
        & \makecell[l]{
            %\tabitem Better AIGC models\\
             \tabitem Slow the adoption of AIGC \\
            \tabitem Less creative works
        }
        \\
        \midrule
        
        Model User 
        & \makecell[l]{
              \tabitem More motivation to explore AIGC untapped potential\\
                \tabitem More creative works 
        }      
        & \makecell[l]{
                \tabitem Rely on AIGC models\\ 
                \tabitem Hard to judge the originality
        }      
        % \makecell 命令中不可有空行
        & \makecell[l]{
                \tabitem Slow the adoption of AIGC\\
                   \tabitem Huge overhead on the Copyright registration system \\due to massive generated content 
        }
        \\
        \bottomrule
    \end{tabular}
    }
    \caption{A summary of the what-if analysis: there is no ideal solution to solve the AIGC copyright issue, which causes the AIGC copyright dilemma. Thus, we need an option that could \textbf{alleviate} the dilemma. }
    \label{tab:1}
\end{table*}

\section{The AIGC Copyright Dilemma: A What-if Analysis}
In this section, we leverage the what-if analysis to clarify the AIGC copyright dilemma. The what-if analysis method~\citep{zagonel2004using,kleijnen1994sensitivity} is a decision-making process that is widely used in evaluating complicated scenarios and potential risks beforehand, thereby helping make the right decision. Following are the three scenarios we consider in terms of  \textbf{benefits}, \textbf{challenges}, and \textbf{risks}, respectively. 

\subsection{What if the data owner claims the copyright?}

%\subsubsection{Benefits} 
\noindent \textbf{Benefits:} This way could best protect the copyright of the training data, which also complies with the copyright law to protect the creation in the public domain. This practice has been used in non-AIGC domains since the debut of the copyright law. For example, an artist creates an artwork in a physical format and posts the electronic copy online. Whether it is online or offline, the artist could claim the copyright of the artwork and legally share the revenue from the artwork. Such practice has boosted the sustainability of corresponding industry sectors with the prosperous development of the economy.

%Analogy to the AIGC field, like those who develop the technology to convert physical artworks into digital forms and provide related technical support, the model owner transforms raw data into a powerful model by learning virtual logic from the training set.Model users are those who create secondary works based on the original work.The critical consideration in determining whether a secondary creation is infringing is originality. If AIGC cannot prove its originality and cannot be significantly different from the original work, its copyright should still belong to the data owner.

%In addition, if the data owner claims the AIGC copyright, the copyright involved in the lifecycle of AIGC, whether the training set or AIGC, is only related to the data owner.Model users no longer face the risk of infringement by claiming AIGC's copyright, which has already been generated by others and protected by copyright.\\

%\subsubsection{Challenges} 
\noindent \textbf{Challenges:} Unlike traditional proprietary materials, the training data of AIGC models is mostly from the online content of the Internet\citep{commonCrawl}, along with some private and high-quality data already in hand. 
It has the following properties: 1) the scale is huge, which covers most of the public content online; 2) the data sources are very diverse, which varies from high-quality collections to low-quality content of free riders; 3) the number of data owners is hard to count, and the identities of these data owners are impossible to trace.  It is extremely challenging to identify the exact owner of every piece of the massive training data. The consent from every data owner is almost impossible to achieve in practice.   

%It is difficult to determine whether the training set of the generated model contains copyrighted data, at least not with the existing data traceability technology\citep{wang2023}. Even if infringement caused by generative models can be detected, specific quantitative penalties may require technical support from the currently developing field of data pricing\citep{pei2020survey}. \\

%\subsubsection{Risks} 
\noindent \textbf{Risks:} We assume we could identify all the data owners. Then, if the data owners claim to share the copyright of the AIGC content, there is no way to quantify the attribution, which could be a mess. The reality that we currently cannot identify the data owners makes it even worse. Any data owner who resorts to claiming the copyright by suing the AIGC model owners may suspend the development of AIGC applications and, in return, harm the booming AI industry. 

%If the data owner claims the AIGC copyright, even though model users generate AIGC with prompts constructed by themselves, they must obtain authorization from the data owner before use.This will discourage the user and is not conducive to the development of the user ecology.

\subsection{What if the model owner claims the copyright?}
%\subsubsection{Benefits} 
\noindent \textbf{Benefits:} It is an expensive process to create a workable AIGC model, which involves not only expensive computational costs but also the expensive cost of talented scientists and engineers~\citep{strubell-etal-2019-energy,sharir2020cost}.
With the vast amount of investment, model owners expect to claim the AIGC copyright to protect themself and seek more revenue in this business sector, which could, in return, benefit the economy. 
This will allow model owners to have more funds for AIGC model research and training, thereby obtaining better-performing models, promoting the development of AIGC, and providing better services to users.

%Moreover, there are already some watermarking technologies \citep{kirchenbauer2023watermark} which could be used to identify the model owner of generated content. This makes it technically feasible for the model owner to claim the copyright.\\

%\subsubsection{Challenges} 
\noindent \textbf{Challenges:} The copyright claim of the model owner will face two types of contribution disputes, both of which may lead to the model owner being sued for infringement.
First, crawling the public online Internet content without data owners' consent to train the AIGC model; this congenital inadequacy is like a chain to prevent model providers from moving forward to claim the copyright.
Second, for model users who provide prompts, it is difficult to accept that they spend money to create copyrights for others.

%\subsubsection{Risks}      
\noindent \textbf{Risks:} If model owners claim the copyright of the AIGC content, it would have two direct risks. On the one hand, this would limit the spreading speed of the AIGC models and become an obstacle to accelerating their development, namely, shooting themself in the foot. On the other hand, people are more hesitant to adopt AIGC in their daily lives, which deviates from the original purpose of inventing AIGC -- raising industry productivity that brings more benefits to society. Both consequences are harmful to model owners.

\subsection{What if the model user claims the copyright?}

%\subsubsection{Benefits} 
\noindent \textbf{Benefits:} Though AIGC is powerful enough to generate content, it still has a gap to reach the ideal content. To bridge this gap, model users spend efforts and expertise to finetune the prompts\citep{chen2023unleashing}. Such efforts are protected so that more users are motivated to explore the AIGC's untapped potential.

%\subsubsection{Challenges}

\noindent \textbf{Challenges:} It is challenging for model users to claim the copyright of AIGC content as they still rely on the AIGC model. The only input they have is the prompt. Since the AIGC content results from the Human-AI collaboration, it is impossible to claim the copyright by excluding the model part. 
Moreover, the creation behavior of model users using AIGC models can be regarded as re-creations or secondary creations to a certain extent, and there is still controversy over whether secondary creations have copyright\citep{wong2008tranformative,lan2023innovation}.

%\subsubsection{Risks} 
\noindent \textbf{Risks:} If the model user claims the copyright of AIGC content, it has a direct impact on slowing the spreading of AIGC. This is because other users may obtain the same or similar content out of the AIGC model but are not aware of its copyright. To avoid further legal risks, people hesitate to use the AIGC in their work, which again deviates from the original purpose of inventing AIGC to raise productivity. Furthermore, due to the fast speed of generating content from AIGC models, the scale of generated content could put a lot of overhead on the copyright registration system, which hinders society's development. 

\subsection{Summary of What-if Analysis}
Through the what-if analysis above, we can find that there is no ideal solution that could take into account the interests of every stakeholder. Every stakeholder has his own grounds to appeal for the copyright interest. Most importantly, AIGC is created via the synergy of the data owner, model owner, and model users, any stakeholder alone could not develop AIGC at all. It is challenging for current regulations to address the intricacies of the AIGC copyright dilemma. On the one hand, strictly enforcing current existing copyright regulations may probably halt the blooming of AIGC. On the other hand, if we place no restrictions on the copyright claim, it could also impose legal and management risks that slow the AIGC adoption in other industry sectors. We believe that the most important function of AIGC is to facilitate social productivity and boost the economy. Therefore, we must balance stakeholders' interests and seek a compromise scheme, which is also the motivation for the advocate of copyleft in previous work.

\section{Case Study Under the Copyleft}
As mentioned in the introduction, there are three real-world cases of AIGC copyright disputes. 

\subsubsection{Case 1:
The New York Times sued OpenAI}
When the copyleft license is applied to the generated content, this contradiction could be alleviated. 
As the model owner, OpenAI does not own the copyright of generated content and cannot profit from the copyright. This will significantly limit the commercial use of generated content and reduce competition with the New York Times, thereby alleviating the dispute.

% On the other hand, the New York Times is worried that OpenAI may steal its own users.
% There are many efforts to prevent large models from outputting their content intact, and through technical means \todo{add citation}, infringements at this point can be detected and reduced. News has timeliness that large models do not have, and this problem will gradually be alleviated.

\subsubsection{Case 2:
Unsupported AIGC copyright claim}
According to our proposed method, the generated content is licensed under a copyleft license. 
Therefore, the model users cannot claim the copyright of the images alone, which directly resolves the dispute. 

\subsubsection{Case 3: Supported AIGC copyright claim}
If the content generated by the plaintiff is licensed under a copyleft license, then the defendant's use of the generated content must be under the copyleft license. Otherwise, it is illegal, which quickly resolves the dispute.

% \subsection{Permissive license and Copyleft}
% Why copyleft? why not other open-soure license?
% (As a candidate, add it if the content is not enought)

%\subsection{One Major Limitation of Copyleft}
%Copyleft can indeed alleviate conflicts among stakeholders and promote the development of AIGC. However, there is still one major limitation of AIGC under the copyleft: copyright protection for data owners remains limited. Infringement of the data owner's copyright is passed through the model to the generated content. Copyleft prevents others from profiting by privatizing the generated content, but for the data owner, the infringement remains. This is a conflict between model owners and data owners, and we think there will be better solutions as infringement detection\citep{yu2023codeipprompt,wang2023} and data pricing\citep{pei2020survey} evolve.
    
% \end{enumerate}

\section{Public Perception: A Survey Method}
We developed a survey to uncover public opinions on copylefting AIGC. It roughly consists of two parts: the AIGC copyright dilemma and the feasibility of copyleft as an alleviation. 22 items were designed, 8 of which were displayed on an ordinal scale of 1 to 5, 8 were yes-or-no questions, and the remaining 6 items were on nominal scales ranging from 4 to 11. 
The detailed questionnaire contents can be found in Appendix A of the supplementary.
A total of 250 participants were recruited through an IRB-approved process. After removing participants who did not pass the attention check, we had a total of 211 of them providing valid responses. 

\subsection{Results}

\subsubsection{The AIGC Copyright Dilemma}
\begin{figure*}[ht]
\centering
\begin{minipage}[t]{0.36\linewidth}
\setlength{\abovecaptionskip}{0.1cm} 
\includegraphics[width=\linewidth]{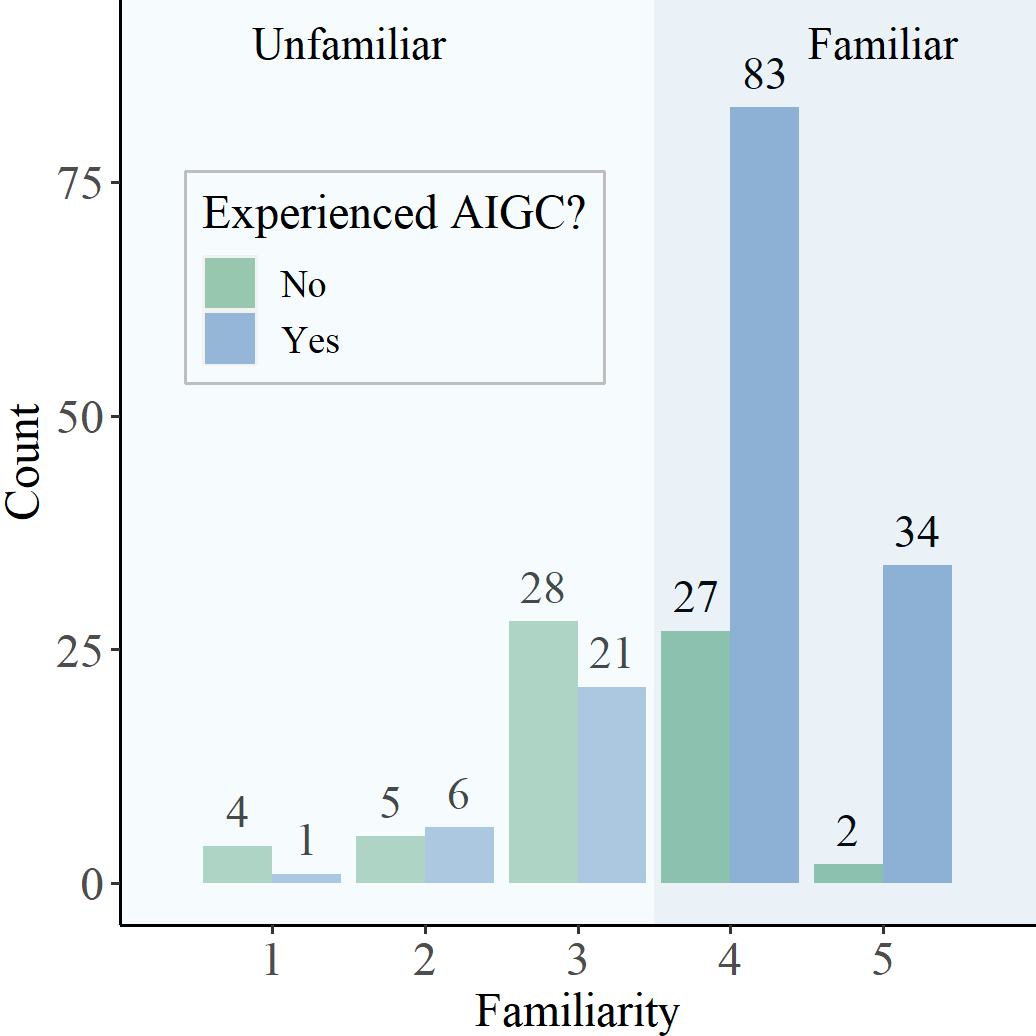}
  \caption*{Figure 1: Familiarity with copyright across whether experienced AIGC. 1 = Never heard, 2 = Heard before, 3 = Limited knowledge, 4 = Familiar, and 5 = Very familiar.}
  \minilabel{localref}\label{fig:2_a}
\end{minipage}
\hfill
\begin{minipage}[t]{0.36\linewidth}
\setlength{\abovecaptionskip}{0.1cm} 
\includegraphics[width=\linewidth]{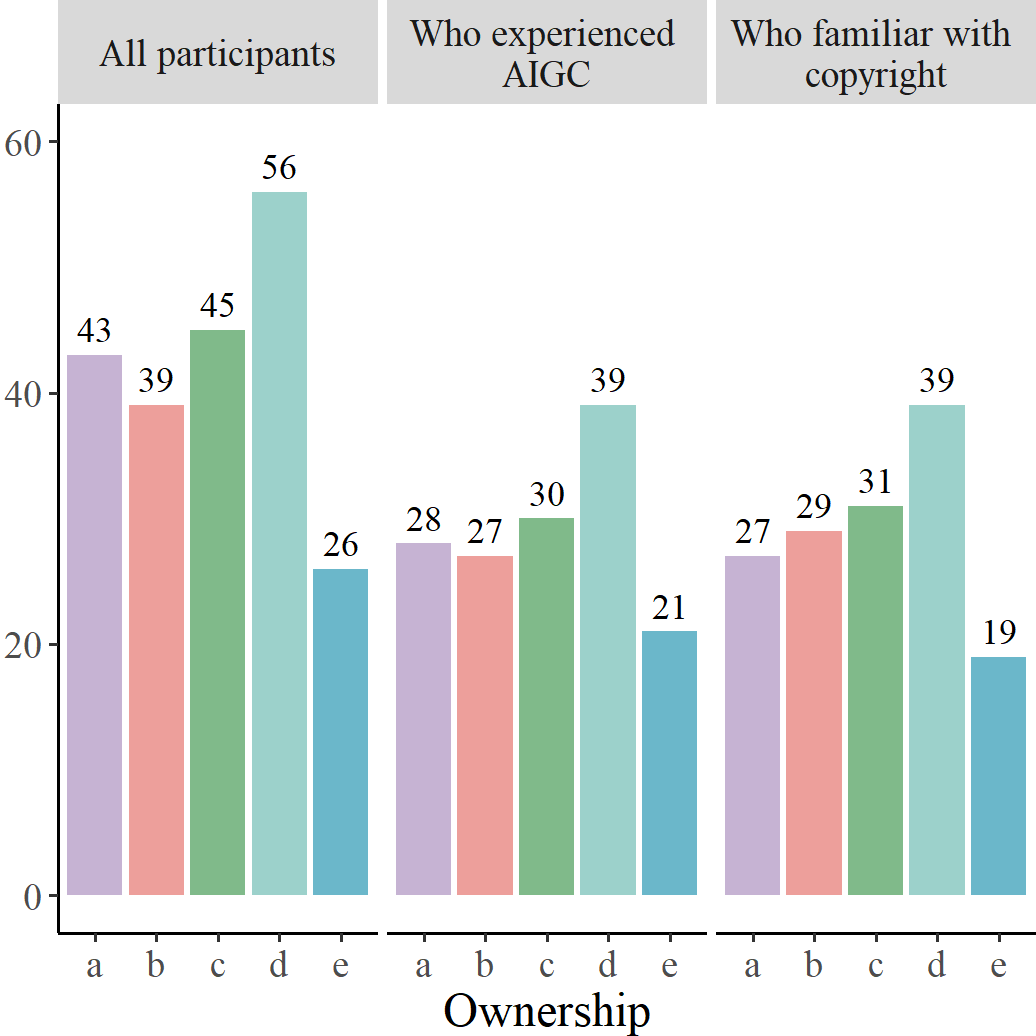}
      \caption*{Figure 2: Copyright ownership of an AI-generated image across different groups (all participants, participants experienced AIGC, and participants familiar with copyright). a = van Gogh, b = da Vinci, c = Model user, d = Model owner, and e = Data owner.}
      \minilabel{localref}\label{fig:2_b}
\end{minipage}
\hfill
\begin{minipage}[t]{0.25\linewidth}
\setlength{\abovecaptionskip}{0.1cm} 
\includegraphics[height=0.28\textheight]
{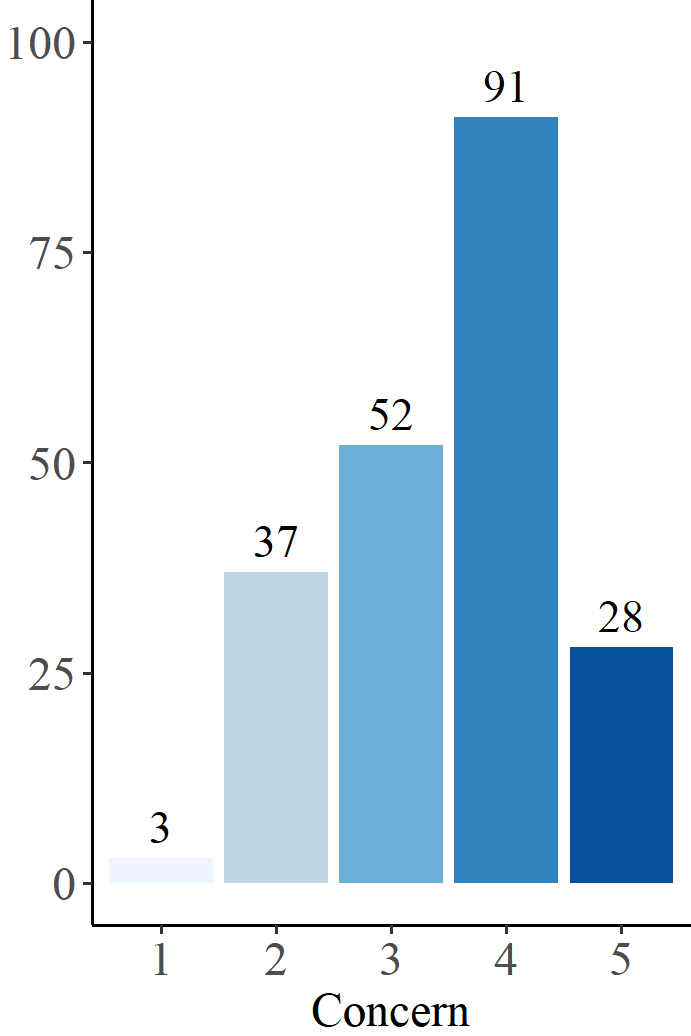}
      \caption*{Figure 3: Concern about AIGC copyright issue, 1 = completely indifferent, 2 = indifferent, 3 = neutral, 4 = Concerned, and 5 = Very concerned.}
      \minilabel{localref}\label{fig:2_c}
\end{minipage}
% \caption*{}
% \label{fig:2}
\vspace{-1.0em}
\end{figure*}

As shown in Figure \ref{fig:2_a}, the majority of our participants have experienced AIGC models and have at least limited knowledge about copyright. However, exposure to AIGC and familiarity with copyright were not enough to lift people out of the AIGC copyright dilemma.

% % 2,3,4
% \begin{figure*}[ht]
% \centering
% \subfloat[Familiarity with copyright across whether experienced AIGC. 1 = Never heard, 2 = Heard before, 3 = Limited knowledge, 4 = Familiar, and 5 = Very familiar.]{
%   \includegraphics[width=0.35\linewidth]{survey_images/fig2_1.png}
%   \label{fig:2_a}
% }
% \hfill
% \subfloat[Copyright ownership of an AI-generated image across different groups (all participants, participants experienced AIGC, and participants familiar with copyright). 1 = van Gogh, 2 = da Vinci, 3 = Model user, 4 = Model owner, and 5 = Data owner.]{
%       \includegraphics[width=0.35\linewidth]{survey_images/fig3_1.png}
%       \label{fig:2_b}
% }
% \hfill
% \subfloat[Concern about AIGC copyright issue, 1 = completely indifferent, 2 = indifferent, 3 = neutral, 4 = concern about, and 5 = strongly concern about.]{
%       \includegraphics[height=0.27\textheight]{survey_images/fig4_1.png}
%       \label{fig:2_c}
% }
% \caption{}
% \label{fig:2}
% \end{figure*}

% 2,3,4

% % 5,6,7
% \begin{figure*}[ht]
% \centering
% \subfloat[Concern about AIGC copyright issue across different groups (``experienced AIGC: no" VS ``experienced AIGC: yes" for the left panel, and ``unfamiliar with copyright" VS ``familiar with copyright" for the right panel). The x-axis has a scale identical to Figure \ref{fig:2_c}.]{
%     \includegraphics[width=0.32\linewidth]{survey_images/fig5.png}
%   \label{fig:3_a}
% }
% \hfill
% \subfloat[Relationship between copyleft and copyright. 1 = completely opposed, 2 = opposed, 3 = not sure, 4 = not opposed, and 5 = completely opposed.]{
%   \includegraphics[width=0.3\linewidth]{survey_images/fig6.png}
%   \label{fig:3_b}
% }
% \hfill
% \subfloat[Relationship between copyleft and copyright across different groups (``experienced AIGC: no" VS ``experienced AIGC: yes" for the left panel, and ``unfamiliar with copyright" VS ``familiar with copyright" for the right panel). The x-axis has a scale identical to Figure \ref{fig:3_b}.]{
%       \includegraphics[width=0.32\linewidth]{survey_images/fig7.png}
%   \label{fig:3_c}
% }
% \caption{}
% \label{fig:3}
% \end{figure*}

% 5,6,7
\begin{figure*}[ht]
\centering
\begin{minipage}[t]{0.33\linewidth}
\setlength{\abovecaptionskip}{-0.25cm} 
    \includegraphics[width=\linewidth]{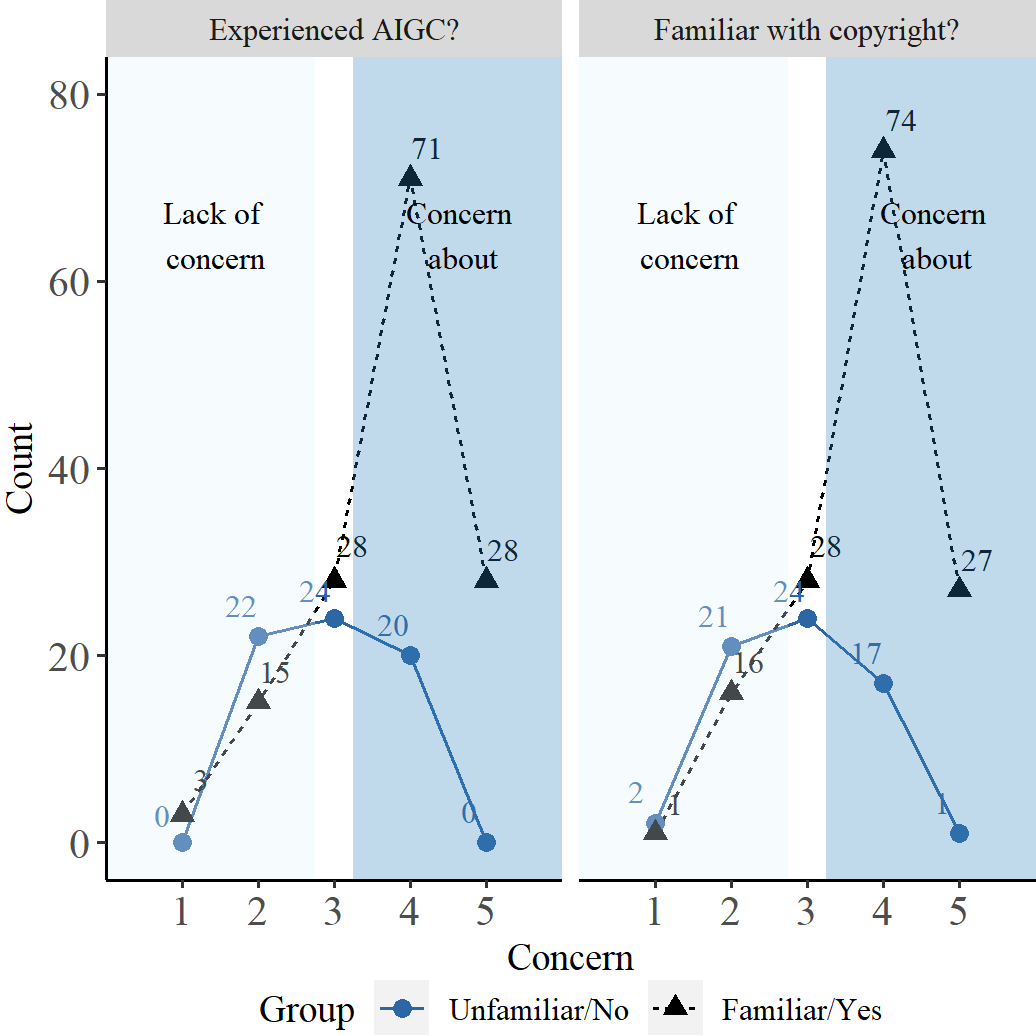}
  \minilabel{localref}\label{fig:3_a}
  \caption*{Figure 4: Concern about AIGC copyright issue across different groups (``experienced AIGC: no" VS ``experienced AIGC: yes" for the left panel, and ``unfamiliar with copyright" VS ``familiar with copyright" for the right panel). The x-axis has a scale identical to Figure \ref{fig:2_c}.}
\end{minipage}
\hfill
\begin{minipage}[t]{0.31\linewidth}
\setlength{\abovecaptionskip}{-0.25cm} 
  \includegraphics[width=\linewidth]{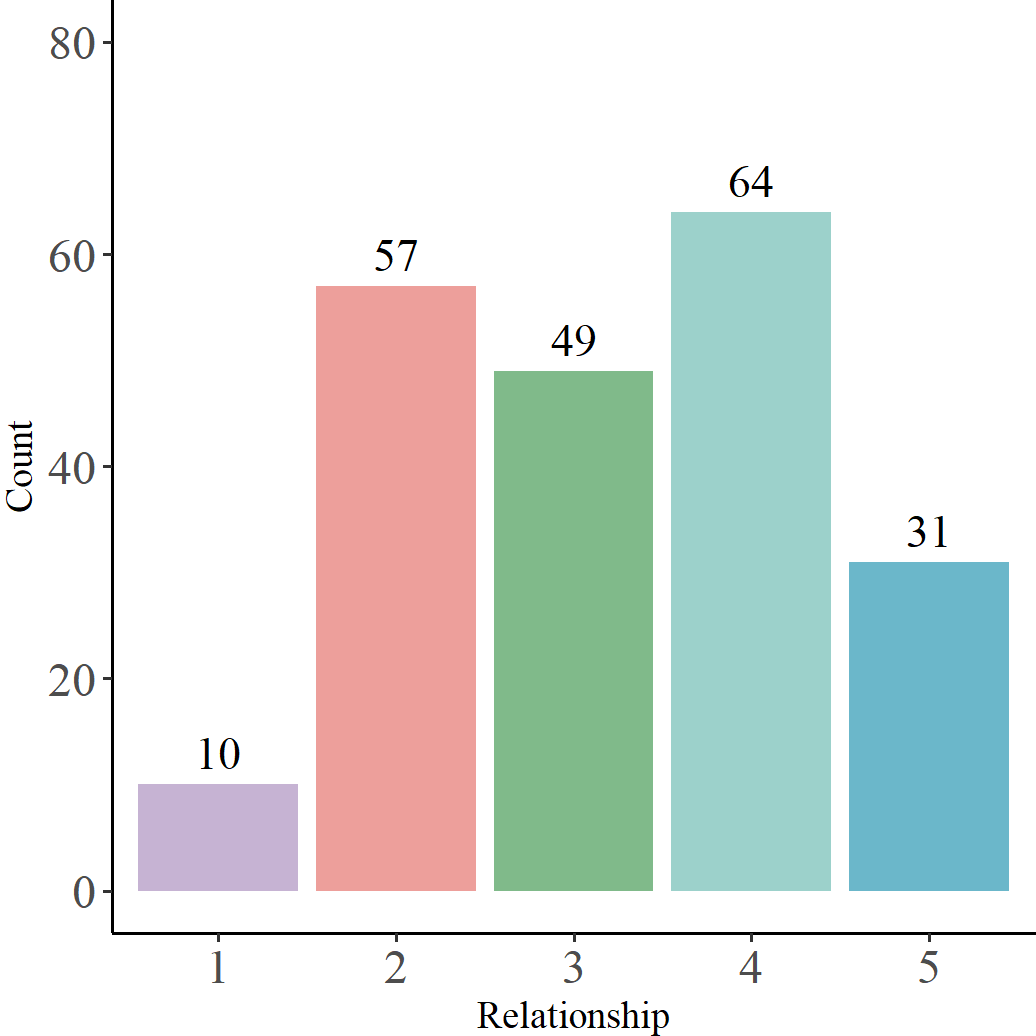}
  \minilabel{localref}\label{fig:3_b}
  \caption*{Figure 5: Relationship between copyleft and copyright. 1 = very conflicted, 2 = somewhat conflicted, 3 = unsure, 4 = somewhat not conflicted, and 5 = Not conflicted at all.}
\end{minipage}
\hfill
\begin{minipage}[t]{0.33\linewidth}
\setlength{\abovecaptionskip}{-0.25cm} 
\includegraphics[width=\linewidth]{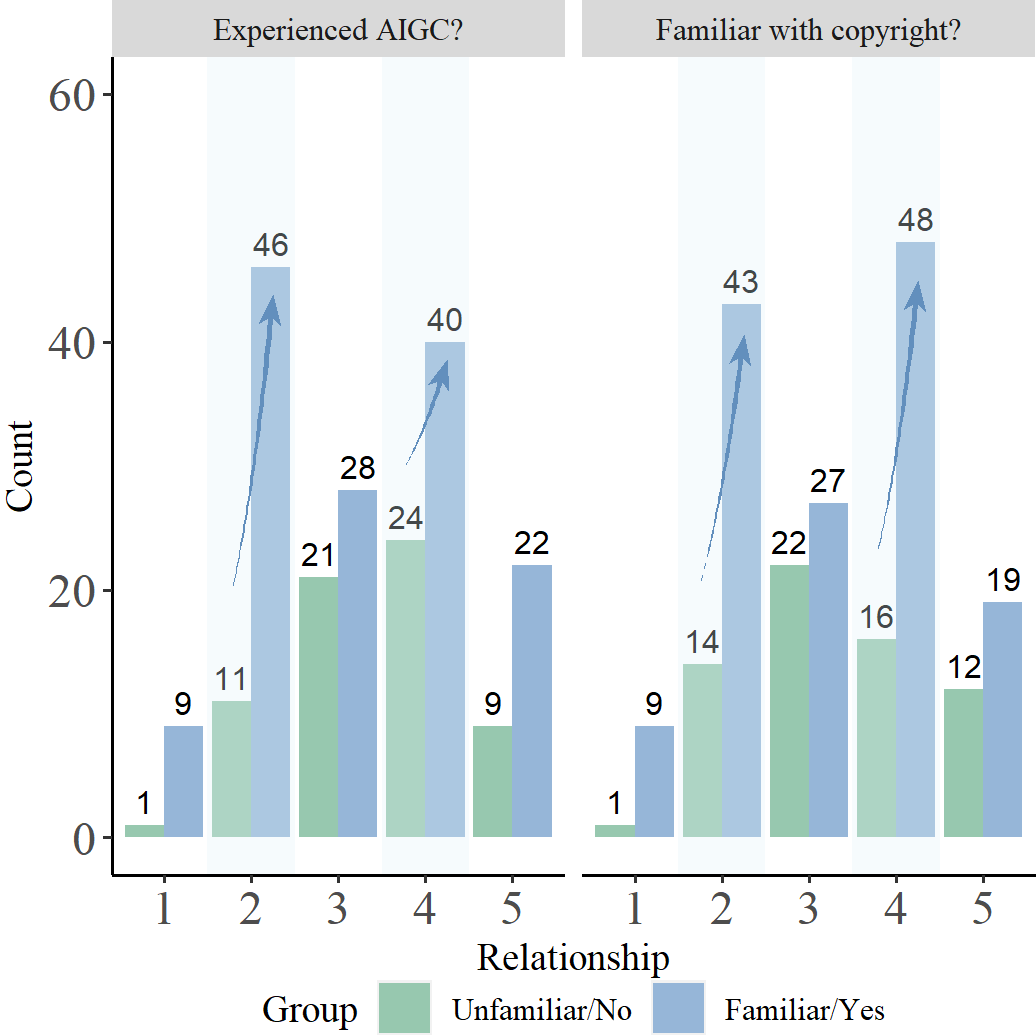}
  \minilabel{localref}\label{fig:3_c}
  \caption*{Figure 6: Relationship between copyleft and copyright across different groups (``experienced AIGC: no" VS ``experienced AIGC: yes" for the left panel, and ``unfamiliar with copyright" VS ``familiar with copyright" for the right panel). The x-axis has a scale identical to Figure \ref{fig:3_b}.}
\end{minipage} 
% \caption{}
%\label{fig:3}
\vspace{-1.0em}
\end{figure*}

% 8,9
% \begin{figure*}[ht]
% \centering
% \subfloat[Level of agreement on using authorized AIGC and extent of freedom expected when using AIGC across different groups (``AIGC copyright infringement after copylefting: no" VS ``AIGC copyright infringement after copylefting: yes").]{
%   \includegraphics[width=0.48\linewidth]{survey_images/fig8.png}
%     \label{fig:4a}
% }
% \hfill
% \subfloat[Number of participants selecting different number of upsides and downsides.]{
%     \includegraphics[width=0.4\linewidth]{survey_images/fig9.png}
%     \label{fig:4b}
% }
% \caption{}
% \label{fig:4}
% \end{figure*}

\begin{figure*}[ht]
\centering
\begin{minipage}[t]{0.48\linewidth}
\setlength{\abovecaptionskip}{-0.cm}
  \includegraphics[width=\linewidth]{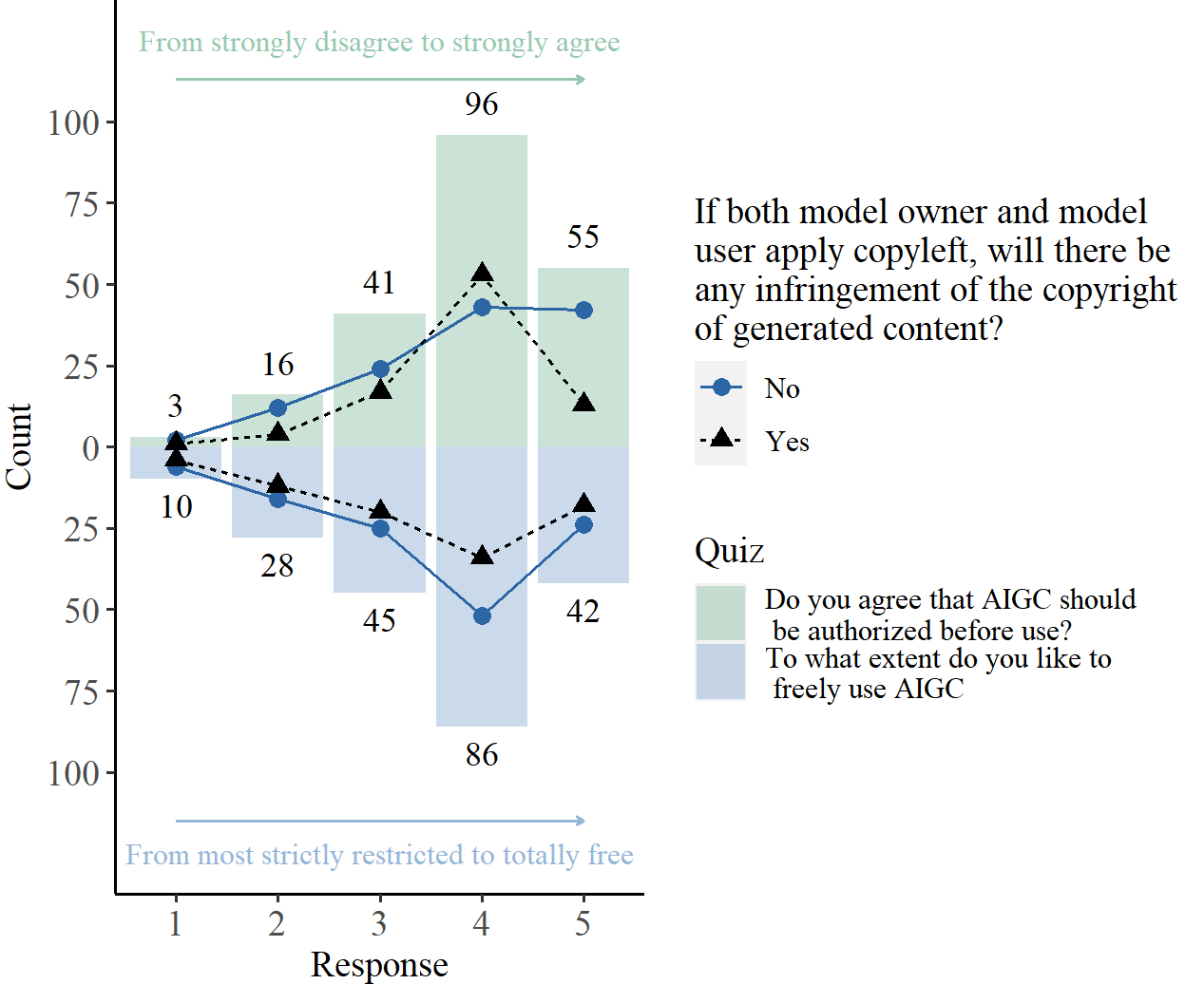}
    \caption*{Figure 7: Level of agreement on using authorized AIGC and extent of freedom expected when using AIGC across different groups (``AIGC copyright infringement after copylefting: no" VS ``AIGC copyright infringement after copylefting: yes").}
    \minilabel{localref}\label{fig:4a}
\end{minipage}
\hfill
\begin{minipage}[t]{0.4\linewidth}
\setlength{\abovecaptionskip}{-0.cm}
    \includegraphics[width=\linewidth]{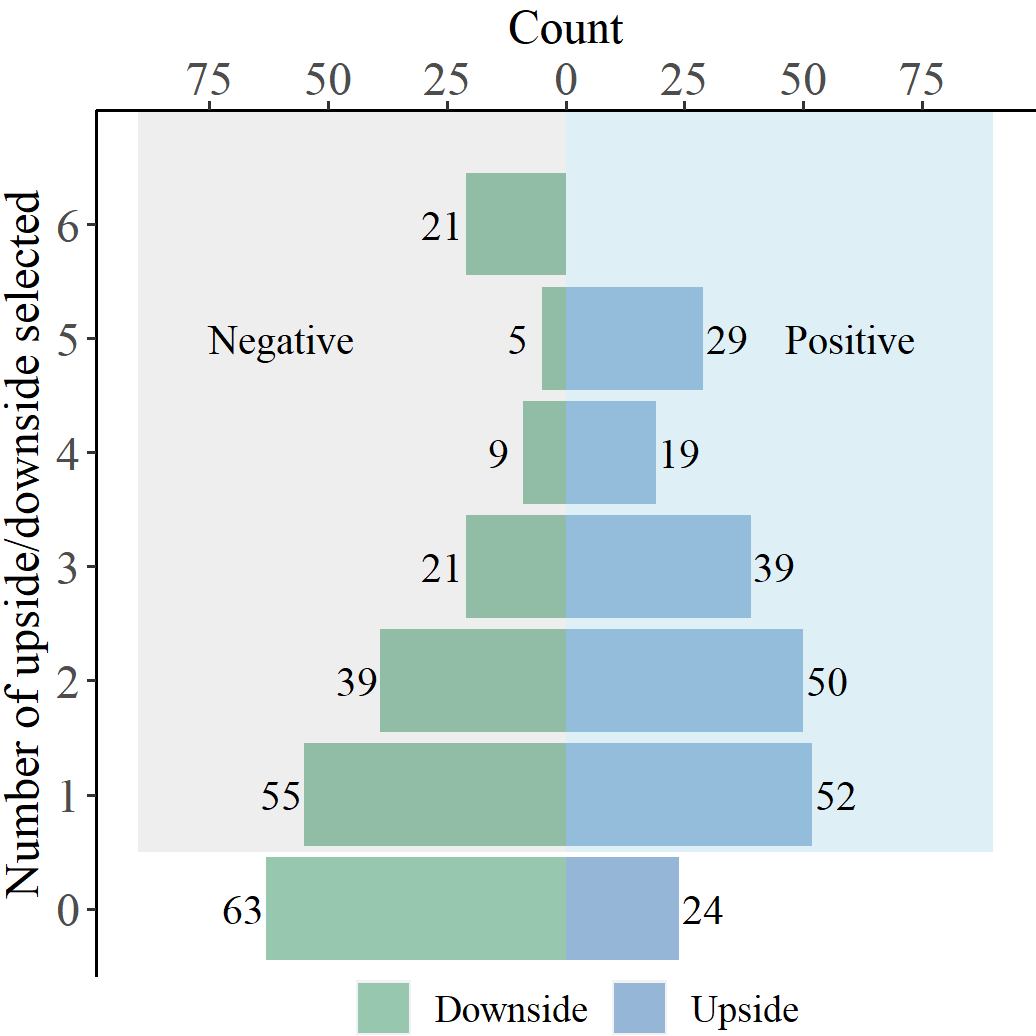}
    \caption*{Figure 8: Number of participants selecting different numbers of upsides and downsides.}
    \minilabel{localref}\label{fig:4b}
\end{minipage}
%\caption{}
%\vspace{-1.0em}
\label{fig:4}
\end{figure*}

% \begin{figure}[ht]
%   \centering
%   \includegraphics[width=\linewidth]{survey_images/familiarity_fill-1.png}
%   \caption{}
%     \label{fig:4}
% \end{figure}

\textbf{People were confused about the ownership of AIGC copyright}. Participants were presented with a fictitious question about who should own the copyright of an AI-generated Mona Lisa in Vincent Van Gogh style. Five options were provided for them to choose from(see Figure \ref{fig:2_b} for more details). It is clear that fewer people considered data owners should own the copyright. We surmise this is because a) the public's intuition that it is hard to copyright data, and b) in AI-related scenarios, the algorithm/model behind it is directly responsible for the content generated and is usually given more emphasis than the data used for training. We, therefore, excluded the ``data owner" option in the following statistical analysis. 

A \(\chi^2\) test revealed that the difference among the numbers of endorsements of the rest four options was statistically insignificant (\(\chi^2=3.47\), \(df=3\), \(p=0.32\)). This pattern was consistently observed among people who were familiar with copyright (\(\chi^2=2.63\), \(df=3\), \(p=0.45\)) and among people who have used AIGC before (\(\chi^2=2.90\), \(df=3\), \(p=0.41\)). Given this prevailing confusion, it was not surprising that more than half of our participants thought it was not easy to establish copyright for AIGC (\(n=121\), \(57.35\%\)). Familiarity with copyright failed to change this perception, for those familiar with copyright, about \(56.46\%\) of them recognized the difficulty, whereas the ratio for people who were unfamiliar with copyright was \(57.58\%\), no statistical difference observed between these two proportions (\(z=-0.12\), \(p=0.91\)). The experience in AIGC had no effect either (\(z=0.11\), \(p=0.92\)).

% \begin{figure}[!ht]
%   \centering
%   \includegraphics[width=1.0\linewidth]{survey_images/onwership-1.png}
%   \caption{Who shall own the copyright of a Mona Lisa in Vincent Van Gogh style by Midjourney.}
%     \label{fig:5}
% \end{figure}

\textbf{People concerned about the AIGC copyright issue.} As shown in Figure \ref{fig:2_c}, \(56.40\%\) participants worried about the copyright of AIGC, \(24.64\%\) of them were neutral, only \(18.96\%\) of them lacked in concern. The number of people who lacked in concern remained stable across the two groups (see the left part of each panel of Figure \ref{fig:3_a}). But the number of people who were concerned about the issue varied dramatically in groups; the AIGC copyright issue was more likely to be on the radar of people who were familiar with copyright or used to be AIGC users (see the right part of each panel of Figure \ref{fig:3_a}).

% Added trim to crop the white space above and below the image. by gxw
% \begin{figure}[!ht]
%   \centering
%   \includegraphics[width=\linewidth, trim=0cm 1.0cm 0cm 1.0cm, clip]{survey_images/concern_all-1.png}
%   \caption{xxx.}
%     \label{fig:6}
% \end{figure}

% \begin{figure}[!ht]
%   \centering
%   \includegraphics[width=\linewidth]{survey_images/concern_facet-1.pdf}
%   \caption{Concern about copyright issue in AIGC.}
%     \label{fig:7}
% \end{figure}

\textbf{People were unclear about the relationship between copyleft and copyright.} As shown in Figure \ref{fig:3_b}, less than half of our participants thought that copyleft is not the antithesis of copyright (\(n=95\), \(45\%\)), which is the correct understanding of the relationship between copyleft and copyright. We further examined the pattern by conducting a \(\chi^2\) test (see Figure \ref{fig:3_c} for more details). When doing so, we excluded the ``very conflicted" option because the cell size was smaller than five in both the group of participants who were unfamiliar with copyright and the group of participants with no experience in AIGC. The results suggested that whether people were familiar with copyright tended to influence their attitude towards the relationship in interest (\(\chi^2=7.30\), \(df=3\), \(p=0.06\)). Surprisingly, it seemed that the knowledge of copyright and experience in AIGC tended to shape the response of our participants into a bimodal distribution. In particular, among participants who had prior knowledge of copyright, there was a \(153.85\%\) increase in the number of individuals who accurately perceived the relationship as ``somewhat not conflicted" compared to those who were unfamiliar with copyright. the number of people who were familiar with copyright but chose the ``somewhat conflicted" category also increased by \(66.67\%\). A similar pattern was clearly shown in the right panel of Figure \ref{fig:3_c}. Familiarity with copyright and experience in AIGC seemed to complicate the problem.

% \begin{figure}[!ht]
%   \centering
%   \includegraphics[width= 0.8\linewidth]{survey_images/relationship-1.pdf}
%   \caption{Relationship between copyleft and copyright.}
%     \label{fig:8}
% \end{figure}
% 
% 
% \begin{figure}[!ht]
%   \centering
%   \includegraphics[width=\linewidth]{survey_images/relationship_facet-1.pdf}
%   \caption{Relationship between copyleft and copyright.}
%   \label{fig:9}
% \end{figure}

Immediately after the question about AI-generated images, we asked our participants a yes-no question about whether they think they can use the generated picture freely. No statistically significant difference between the number of people who selected ``yes" and ``no" (\(\chi^2=1.37\), \(df=1\), \(p=0.2419\)), which further demonstrated the ambivalence rooted in the AIGC copyright dilemma.

\subsubsection{Copyleft is a feasible option}

\textbf{People inclined to use authorized AIGC under restrictions.} Three questions were presented to the participants about their expected level of authorization and freedom when using AIGC, the results were captured in Figure \ref{fig:4a}. Clearly, we observed two negatively skewed distributions of bar, implying that more people a) supported obtaining authorization of AIGC before use (\(n=96+55\), \(71.56\%\)), and b) preferred to be imposed on loose restrictions or have no restriction at all while using AIGC (\(n=86+42\), \(60.66\%\)). More specifically, the option we considered as most reasonable, using authorized AIGC under loose restrictions (response = 4 in Figure \ref{fig:4a}), received the strongest endorsement from our participants.

\textbf{People generally agreed that copyleft could resolve AIGC copyright infringement.} Meanwhile, more than half of our participants (\(n=123\), \(58.29\%\)) thought copyleft is capable of preventing copyright infringement of AIGC if properly used by both model owner and model user. Interestingly, the largest difference between the number of participants who chose ``no" (\(n=42\)) and ``yes" (\(n=13\)) was observed in the bar at the upper-right corner of Figure \ref{fig:4a}. That is, among the people who strongly agreed in obtaining authorization of AIGC (\(n=55\)), \(79.25\%\) of them thought copylefting AIGC could prevent copyright infringement.

% \begin{figure}[ht]
%   \centering
%   \includegraphics[width=\linewidth]{survey_images/fig8.png}
%   \caption{Level of agreement on using authorized AIGC and extent of freedom expected when using AIGC across different groups (``AIGC copyright infringement after copylefting: no" VS ``AIGC copyright infringement after copylefting: yes").}
%     \label{fig:10}
% \end{figure}

\textbf{People were generally positive about applying copyleft to AIGC.} We asked participants what will happen if AIGC is copylefted. Six negative predictions and five positive predictions were provided; they were free to choose any number of these upsides and downsides. As shown in Figure \ref{fig:4b}, people were generally positive about the future of applying copyleft to AIGC in that a) the number of participants who selected at least one downside (\(n=150\)) was smaller than the number of participants who selected at least one upside (\(n=185\)), and b) \(73.71\%\) participants opted for no more than two downsides and \(65.55\%\) participants selected two or more upsides.

% \begin{figure}[ht]
%   \centering
%   \includegraphics[width=0.8\linewidth]{survey_images/fig9.png}
%   \caption{Number of participants selecting different number of upsides and downsides.}
%     \label{fig:11}
% \end{figure}

\subsubsection{Willingness to utilize copyleft in AIGC}
We used an ordinal logistic regression model with \textit{Willingness to Use Copyleft in AIGC} as the numeric dependent variable (1-5). Ten variables investigated in the two sub-sections above were included as independent variables (see Table \ref{tab:3} for more details). The results consist of two parts: AIGC copyright dilemma and Copyleft as a feasible option.

\textbf{AIGC copyright dilemma: confusion and difficulty over assessing the relationship between copyleft and copyright decreased the will, but concern about copyright issue did the opposite}. \textit{Familiarity with copyright} (\(\beta=-0.01\), \(p=0.97\)) and \textit{Experience in AIGC} (\(\beta=0.23\), \(p=0.49\)) have no measurable effect on \textit{Willingness to Use Copyleft in AIGC}. \textit{Relationship Between Copyleft and copyright}, and \textit{Difficulty in Establishing AIGC Copyright}, both significantly and negatively impacted \textit{Willingness to Use Copyleft in AIGC}. More specifically, the odds decreased by \(100\%-74.43\%=25.57\%\) for every one unit increase in \textit{Relationship Between Copyleft and copyright} (\(\beta=-0.30\), \(p=0.04\)), implying that participants who tended to perceive the relationship between copyleft and copyright as not opposed to each other were less willing to apply copyleft to AIGC. Similarly, the more difficult people felt about establishing copyright for AIGC, the less likely they will accept copyleft (\(\beta=-0.27\), \(p=0.08\), odds reduced by \(23.75\%\)). Whereas \textit{Concern About AIGC Copyright Issue} significantly increased the odds of a participant having a higher willingness to embrace copyleft (\(\beta=0.57\), \(p=0.00\)) by \(77.15\%\).

\textbf{Copyleft as a feasible option: the preference for using authorized AIGC while under loose restrictions increased the will, and so did the number of upsides selected}. \textit{AIGC Copyright Infringement After Copylefting} and \textit{Downsides of copylefting AIGC} failed to have a significant impact on \textit{Willingness to Use Copyleft in AIGC}. \textit{Authorizing AIGC Before Use} had a favorable influence on \textit{Willingness to Use Copyleft in AIGC} (\(\beta=0.73\), \(p=0.00\)), increases the odds of interest by \(107.63\%\) per unit. People expecting a higher level of freedom in using AIGC were more likely to advocate the use of copyleft (\(\beta=0.58\), \(p=0.00\), odds increased by \(78.48\%\)). It seemed that the number of upsides people recognized regarding copylefting AIGC reaped a positive effect on \textit{Willingness to Use Copyleft in AIGC} (\(\beta=0.19\), \(p=0.05\)), the odds increased per unit change was \(21.06\%\).

\begin{table*}[ht]
    \centering
    \resizebox{\linewidth}{!}{
    \begin{tabular}{llllllll}
        \toprule
        & Independent variable   & Coding & \(\beta\)	& \(se\) & \(t\) & \(p\) & Odds ratio(\%) \\
        \midrule
        \cmidrule{1-1} 
        \multirow{5}*{\makecell[l]{Part 1:\\ AIGC copyright dilemma}} &
        \textit{Familiarity with Copyright} & 1-5 &	-0.01& 	0.20&  -0.03& 	0.97	&  \\%0.994\\
        ~ & \textit{Experience in AIGC}& 0-1 &0.23&	0.32&	0.70&	0.49&	\\%1.252\\
        ~ & \textit{Relationship Between Copyleft and Copyright} & 1-5 &	-0.30& 	0.14& 	-2.09& 	0.04& 	$\downarrow$ 74.43\%\\
        ~ & \textit{Difficulty in Establishing AIGC Copyright}&	1-5 & -0.27&		0.15&		-1.77&		0.08&	$\downarrow$ 76.25\%\\
        ~ & \textit{Concern about AIGC Copyright Issue} & 1-5 & 0.57&		0.18&		3.24&		0.00&		$\uparrow$ 177.15\%\\
        \midrule
        \cmidrule{1-1} 
        \multirow{5}*{\makecell[l]{Part 2:\\ Copyleft as a feasible option}} & \textit{Authorizing AIGC Before Use} &	1-5&	0.73&		0.17&		4.38&		0.00&	$\uparrow$ 207.63\%\\
        ~ & \textit{Freedom of Using AIGC} &	1-5&	0.58&		0.14&		4.07&		0.00&	$\uparrow$	178.48\%\\
        ~ & \textit{AIGC Copyright Infringement After Copylefting}&	0-1 &	-0.13&		0.29&		-0.43&		0.67&		\\%0.883\\
        ~ & \textit{Upsides of Copylefting AIGC} &	0-5&	0.19&		0.10&		1.94&		0.05& $\uparrow$ 121.06\%\\
        ~ & \textit{Downsides of Copylefting AIGC}& 0-6 &		-0.02&		0.08&		-0.28&		0.78&		\\%0.978\\
        \bottomrule
    \end{tabular}
    }
    \caption{The results of the ordinal logistic regression model, demonstrating how various factors impact participants' willingness to use copyleft in AIGC.}
    \label{tab:3}
\end{table*}

\section{Implications and Recommendations}

%Prior work has advocated the use of copyleft in AI-related issues \citep{schmit2023leveraging,lilova2021copyright} but rarely aimed at the AIGC copyright dilemma. We took a step further by conducting a mixed-methods study, combining a formal what-if analysis and a carefully designed survey to explore the feasibility of alleviating the dilemma via copyleft. As the AIGC continues to evolve vastly and impact our society profoundly, we discuss our findings to aid in the prosperity of AIGC while bringing the most benefits to our society.

\textbf{Copyleft can effectively alleviate the AIGC copyright dilemma and facilitate the dissemination of AIGC.} Our qualitative analysis has demonstrated that the implementation of copyleft can help mitigate disputes among stakeholders and provide a solid foundation for adjudicating AIGC infringement cases.
Furthermore, copyleft can promote the sharing of AIGC by requiring derivatives to be licensed under the same terms.

\textbf{The AIGC copyright dilemma was well perceived among people and further catalyzed their willingness to adopt copyleft}. Similar to previous work, we echoed the confusion over the ownership of AIGC copyright~\citep{ijcai2023p803,chen2023challenges}. But unlike prior work, we unveiled that the preknowledge of copyright and AIGC could make people more concerned about the AIGC copyright issue but failed to reduce the perplexity. People's concern over the AIGC copyright issue naturally converted into a positive factor in becoming a copyleft adopter. Surprisingly, the correct understanding of the relationship between copyleft and copyright turned out to be a negative factor, so did the perceived difficulty in establishing AIGC copyright ownership. There is the privatization difficulty of copyleft that is to blame. For example, two participants reported ``feeling privatization depressed" if the content they ``create" via the free use of AIGC model has to be freely available to others, thus they were less willing to use copyleft despite they understood the relationship and appreciated the difficulty.

%he reason we would like to argue was that the widely held misconceptions about copyleft (e.g., copyleft and copyright conflict) prevented people from actively recognizing the feasibility of copyleft and further reduced the possibility of being a future copyleft user. 

\textbf{Copyleft satisfied people's need for AIGC, making individuals positive about its future and increasing their inclination to embrace copyleft.} Our participants preferred to use authorized AIGC with loose restrictions. Copyleft squared with this preference in that if adopted, copyleft grants people the permission to use and distribute content \citep{copyleft_definition} freely. Therefore, people with stronger preferences were more confident that copyleft could prevent AIGC copyright infringement. All these built up into an overall positive prediction from the public about what can happen if copyleft is applied to AIGC and, in turn, greatly increased their will to embrace copyleft.

\textbf{Recommendations.} Since AIGC's copyleft license is shared among data owners, model owners, and model users, the construction of the license should also be completed collaboratively. Furthermore, a considerable amount of effort should be placed into promoting copyleft to make more people aware of its feasibility and dispel the detrimental misconceptions about copyleft. This is arguably an across-the-board joint collaboration to be done on the level of our entire society, including stakeholders, policy designers, educators, etc.

\begin{comment}
\textbf{Stage 1 (Data owner):}
Data owners have the obligation to disclose their data copyrights proactively. This reduces the cost of confirming data ownership for model owners during training set construction and minimizes infringement disputes between them.

\textbf{Stage 2 (Model owner):}
When building a training dataset, the model owner should identify and handle the copyrights of all data owners involved in the dataset as thoroughly as possible.
Once the model training is finished, model owners should create an initial copyleft license that includes declarations from both the data owners and themselves.

\textbf{Stage 3 (Model user):}
After building the prompt and generating AIGC, the user needs to add a copyright statement about the prompt in the license to complete the construction of the copyleft license.

\textbf{Stage 4 (Model owner):}
By default, the model owner provides AIGC service for users. To facilitate subsequent rights protection and evidence collection, the model owner should bind the final copyleft license to AIGC through watermarks or other methods.
\end{comment}

\section{Conclusion}
%\todo{re-write the conclusion}
In conclusion, by adopting a mixed-methods approach, we have demonstrated the feasibility of alleviating the AIGC dilemma via copyleft. From the qualitative aspect, 
we first untangled complex conflicts within the AIGC copyright dilemma through what-if analysis. 
Realizing there is no perfect way to consider all stakeholders' interests, copyleft could be a practical mitigation method, and its feasibility has been demonstrated through case studies. From a quantitative perspective, our survey has found that people concerned about the AIGC copyright dilemma. They preferred to use authorized AIGC under loose restrictions, thus were overall positive about copylefting AIGC, laying a good foundation of promoting copyleft. We have also quantified how various factors impacted people's willingness to apply copyleft to AIGC. In particular, the concern, preference and positiveness mentioned above served as incentives. Overall, we need an across-the-board joint collaboration to make more people aware of the feasibility. 

\clearpage
\section{Ethical Statement}
This study was approved by the Institutional Review Board (IRB).

%\section*{Acknowledgements}

\bibliographystyle{named}
\bibliography{ijcai24}

\appendix

% \newpage % 换栏不换页
\clearpage

\clearpage

\end{document}